\documentstyle[12pt,aps,prb,epsfig]{revtex}
\begin{document}
\title{Perturbation theory for the one-dimensional optical polaron}
\author{P. A. Khomyakov\footnote{homyakov@phys.bsu.unibel.by}}

\address{Department of Theoretical Physics,
Belarussian State University,
Fr.\ Skarina av.\ 4, Minsk 220080, Republic of Belarus}
\maketitle
\begin{abstract}
The one-dimensional optical polaron is treated on
the basis of the perturbation
theory in the weak coupling limit.
A special matrix diagrammatic technique is developed.
It is shown
how to evaluate all terms of the perturbation theory for
the ground-state energy of a polaron to any order by means
of this technique.
The ground-state energy is calculated up to the eighth order of the
perturbation theory.
The effective mass of an electron is obtained
up to the sixth order of the perturbation theory.
The radius of convergence of the obtained series is estimated.
\end{abstract}
\section{Introduction}
 Nowadays there is a stable interest in
low-dimensional structures
\cite{dega0,devr0,su,brun,sahoo,ganb2}. The one-dimensional
polaron problem is relevant in semiconductor physics, where with
state-of-the-art nano-lithography it has become possible
to confine electrons
in one direction (quantum wires) \cite{wire} and in linear conjugated
organic polymer conductors \cite{poly1}.
 The treatment of the polaron problem in quantum dots can be found
in Ref. \cite{brun,sahoo}.

 Now there are a lot of theoretical works, where
the polaron problem is investigated by means of
the perturbation theory (PT)
\cite{smond,smond2,roseler,haga,hohler,peetsm,khom}.
The series of the perturbation theory
is useful for verifying the approximate nonperturbative
methods in the weak-coupling limit
\cite{pines,fey,devr,dega,ganb1}.
 The PT for the $N$ - dimensional
polaron has been developed in Ref. \cite{smond}, where
the perspectives of $1/N$-expansions are discussed.
The technique of $1/N$-expansions has been developed later for the
optical polaron in Ref. \cite{smondN}.
Up to now the
first three terms of
the weak-coupling expansion for the ground-state energy of the bulk polaron
have been calculated \cite{smond2} (see also \cite{roseler,haga,hohler}),
as well as
two terms for the surface polaron energy \cite{smond} and
three terms of the wire polaron energy \cite{ganb2,khom}.
The investigation of the
convergence of the PT series of the bulk polaron can be found
in Ref. \cite{conv}.

 In this paper the matrix diagrammatic technique is developed for
an optical large polaron. This technique permits to evaluate any
term of the PT, in principle.
The ground-state energy of the one-dimensional polaron
is calculated up to the eighth order of the PT by using the developed
technique. The radius of convergence of the PT series is estimated
by means of Cauchy-Hadamar's
criterion with respect to the calculated terms of the series.

In Section II the
matrix diagrammatic technique is developed.
In Section III the obtained results for the ground-state energy
and the effective mass of an electron are given.
The radius of convergence of the PT series is also estimated.

\section{The matrix diagrammatic technique for the polaron problem
in the weak coupling limit}
The Hamiltonian of the one-dimensional optical large polaron
is given by \cite{peetsm,dega}
\begin{equation}\label{ham}
\displaystyle{{H} = \frac{{p}^{2}}{2} +
\sum_{k} \omega_{k} a_{k}^{+} a_{k} +
\sum_{k} \left(V_{k}^{*}\, a_{k}^{+} e^{-i k\, {x} } +
V_{k}\, a_{k} e^{i k\, {x} }
\right) },
\end{equation}
where $\omega_{k}$ is a frequency of a phonon with momentum
$k$, for the optical polaron $\omega_{k}=\omega$ does not depend on $k$
 and $V_{k}=2^{1/4} (\alpha /
L)^{1/2}$; ${p}, {x}$ are the momentum and space
operators of an electron;
$a_{k}^{+},a_{k}$ are the creation and
annihilation operators of a phonon with momentum $k$;
$L$ is a normalized length; $\alpha$ acts as a coupling
constant of the electron - phonon interaction.
Our units are such that $\hbar$, $\omega$ and the electron mass are unity.
Below we shall make usual simplifying assumption, such that the crystal
lattice acts like a dielectric medium. It means that we can replace
a sum $\sum_{k}$ by an integral $ L\int{\rm d}k/2\pi $.

Let us consider the weak coupling limit $\alpha \ll 1$ for
the polaron with the Hamiltonian (\ref{ham}).
After doing Lee-Low-Pines transformation
\begin{eqnarray}
&& \label{lee} {H^{\prime}} = U^{-1} {H} U  ,
\\
&& \vert \Psi \rangle = U \vert \Psi^{\prime} \rangle ,
\\
 && U = \exp{ [ i\,
( P - \sum_{k} k a_{k}^{+} a_{k} ) \, x ] } ,
\end{eqnarray}
where $P$ is a c-number representing the total system
momentum, we obtain the Shr\"odinger equation
for Eq. (\ref{ham}) in the next form
\begin{eqnarray}\label{shr2}
&& ( {H}_{0} + {H}_{1} )
\vert \Psi^{\prime} \rangle = E
\vert \Psi^{\prime} \rangle ,  \nonumber
\\
&& {H}_{0} = \frac{1}{2} (
P - \sum_{k} k a_{k}^{+} a_{k}  )^{2} + \sum_{k} a_{k}^{+} a_{k} ,
\\
&& {H}_{1} = \sum_{k} V_{k} \left( a_{k}^{+} +
a_{k} \right) .  \nonumber
\end{eqnarray}
Let us use the conventional perturbation theory \cite{morse} for
the SE Eq. (\ref{shr2}) in the Fock basis, where ${H}_{0}$ is an
unperturbed Hamiltonian. Thus, if the zero-approximation of the vector
state $\vert \Psi^{\prime} \rangle $
is a vacuum state of a phonon field $\vert 0 \rangle$
then the ground-state energy of ${H}_{0}$ is given by
$E_{0}^{(0)} = \langle 0 \vert {H}_{0} \vert 0
\rangle = P^{2}/2$.
It is easy to verify that all odd terms
of the PT are equal to zero $E_{0}^{(1)} = E_{0}^{(3)} =\ldots =0$. The
second order of the PT \cite{smond} is
\begin{eqnarray}\label{sec}
E_{0}^{(2)}&=& \langle {H}_{1}
\frac{1}{h_{0}} {H}_{1}  \rangle
\nonumber \\ &&
 = \frac{2^{1/2} \alpha}{2 \pi} \int_{-\infty}^{\infty}
\frac{{\rm d}k_{1}}{E_{0}^{(0)} -  \frac{1}{2} (P-k_{1})^{2} - 1  } ,
\end{eqnarray}
where $h_{0} = E_{0}^{(0)} - {H}_{0}$ and $\langle 0\vert \ldots \vert 0
\rangle \rightarrow \langle \ldots \rangle $, this term is defined by one
connected diagram (see Eq1.eps).
\noindent
The thick line corresponds to an electron propagation with momentum
$P-k_{1}$. The thin line corresponds to a propagation of a virtual
phonon with momentum $k_{1}$, a bold point on the thick electron line
corresponds to a vertex, where one phonon either creates or annihilates.
Below we shall give the Feynman rules for the connected
diagrams represented in the matrix form. The quantity of the diagrams
increases in the next orders of the PT: two connected diagrams in the forth
order, ten connected diagrams in the sixth order, seventy-four connected
diagrams in the eighth order and so on. There are also unconnected diagrams.
These diagrams can be evaluated by
differentiating the energy terms, which are the sums of the corresponding
connected diagrams, with respect to $E_{0}^{(0)}$ (see below).
Note that all multidimensional integrals corresponding to the diagrams
are evaluated by the residue theory. It can be seen from
(\ref{rule}). Thus, there is a technical problem to generate and
evaluate all diagrams in the higher orders.

Now we shall show how to build the matrix diagrammatic technique
which permits us to generate all connected
diagrams by means of any modern
system of computer algebra (SCA).
Any connected diagram can be represented in the $n$-th order of
the PT by using
$n/2$ by $(n-1)$ matrix $\vert\vert N \vert\vert$, where $n$ is
an even number. For example, let us consider the energy term of the forth
order ($n=4$). It has the following form  \cite{morse}
\begin{eqnarray}
E_{0}^{(4)} = \langle  {H}_{1}
\left(\frac{1}{{h}_{0}} {H}_{1}\right)^{3}
\rangle
+ \frac{1}{2} \;
\frac{\partial}{ \partial E_{0}^{(0)} } \left[ \langle  {H}_{1}
\frac{1}{{h}_{0}} {H}_{1} \rangle \right]^{2}.
\end{eqnarray}
This term is defined by a sum of two connected and one unconnected graphical
diagrams  \cite{smond} (see Eq2.eps).
These diagrams can be written in the following matrix form
\begin{equation}\label{md}
\begin{array}{l}
\displaystyle{ E_{0}^{(4)} =
\pmatrix{1&1&1\\0&1&0} + \pmatrix{1&1&0\\0&1&1} +
(1) \frac{\partial (1)}{ \partial E_{0}^{(0)} } ,}
\end{array}
\end{equation}
where $i$-th row of $\vert\vert N \vert\vert$ describes a history of
propagation of $i$-th phonon with momentum $k_{i}$ $(i=1,2\ldots n/2)$.
And $j$-th matrix column shows a distribution of phonons after
passing $j$-th vertex, where one phonon either creates or annihilates
$(j = 1,2 \ldots  n-1)$.
The value of $N_{ij} = 1$ or $0$ corresponds to the existence
or absence of $i$-th phonon between $j$-th and $(j+1)$-th vertex
respectively.
The generation of all connected diagrams for the $n$-th order is realized by
selecting $n/2$ by $n-1$ matrixes with respect to the next rules
\begin{eqnarray}\label{cond}
&& N_{ij} = 0 \; {\rm or}\; 1, \quad
\sum_{i = 1}^{ n/2 } N_{ij} \neq 0, \quad \sum_{j = 1}^{n-1} N_{ij} \neq
0,
 \nonumber \\
&& \sum_{i = 1}^{ n/2 } N_{i\, 1} = 1, \quad
\sum_{i = 1}^{ n/2 } N_{i\, n - 1} = 1, \quad
\nonumber \\ &&
\sum_{j = 1}^{n-1} \vert N_{ij+1} - N_{ij} \vert = 1.
\end{eqnarray}
We have to keep only one arbitrary matrix among matrixes
which are transformed into each other by permutating
matrix rows.
Thus, the whole set of
connected diagrams can be got in the matrix form by means of any SCA.
Using the graphical diagrammatic technique
\cite{smond} it is easy to
find the next rule for our matrix diagrammatic technique
\begin{eqnarray}\label{rule}
&& \vert\vert  N \vert\vert  \longleftrightarrow
\left(\frac{2^{1/2} \alpha}{2 \pi}\right)^{n/2} \,
\int_{- \infty}^{+ \infty} {\rm d} k_{1}
 \ldots \int_{- \infty}^{+ \infty}
{\rm d} k_{n/2}
\nonumber \\
& \times & \prod_{j = 1}^{n-1} \left[  E_{0}^{(0)} - \frac{1}{2} (
P - \sum_{i = 1}^{ n/2 } N_{ij} k_{i} )^{2} - \sum_{i = 1}^{n/2} N_{ij}
\right]^{-1} .
\end{eqnarray}
Any diagram represented in the matrix form corresponds to an
analytical expression. So that
in accordance with the rule (\ref{rule}) we have for (\ref{md})
$$
\begin{array}{l}
\displaystyle{ (1) \longleftrightarrow
 \frac{2^{1/2}\alpha}{2 \pi}
\int_{-\infty}^{\infty} {\rm d} k_{1}\;
\left[ E_{0}^{(0)} -  \frac{1}{2} ( P-k_{1} )^{2} - 1 \right]^{-1} ,}
\\
\displaystyle{ \pmatrix{1&1&1\\0&1&0} \longleftrightarrow
\frac{2 \alpha^{2}}{ (2 \pi)^{2} }
\int_{-\infty}^{\infty} {\rm d} k_{1} \int_{-\infty}^{\infty} {\rm d} k_{2}\;
\left[ E_{0}^{(0)} -  \frac{1}{2} ( P-k_{1} )^{2} - 1 \right]^{-1} }
\\
\qquad\qquad
\displaystyle{\times
\left[ E_{0}^{(0)} - \frac{1}{2} ( P-k_{1}-k_{2}^{} )^{2} - 2 \right]^{-1}
\; \left[ E_{0}^{(0)} - \frac{1}{2} ( P-k_{1} )^{2} - 1 \right]^{-1} ,}
\\
\displaystyle{ \pmatrix{1&1&0\\0&1&1} \longleftrightarrow
\frac{2 \alpha^{2}}{(2\pi)^{2}} \int_{-\infty}^{\infty} {\rm d} k_{1}
\int_{-\infty}^{\infty} {\rm d} k_{2} \;
\left[ E_{0}^{(0)} -  \frac{1}{2} ( P-k_{1} )^{2} - 1 \right]^{-1} }
\\
\qquad\qquad
\displaystyle{\times
\left[ E_{0}^{(0)} - \frac{1}{2} (P-k_{1}-k_{2}^{})^{2} - 2 \right]^{-1}
\; \left[ E_{0}^{(0)} - \frac{1}{2} (P-k_{2})^{2} - 1 \right]^{-1} .}
\end{array}
$$
In order to summarize all unconnected diagrams in the
$n$-th order
we can use the general structure of the conventional perturbation
theory series
\cite{morse}. Note that
the term of the PT
\begin{eqnarray}
\displaystyle{ E_{0}^{(n) {\rm c} } = \langle {H}_{1}
\left( \frac{1}{{h}_{0}} {H}_{1} \right)^{n-1}
\rangle,  }
\end{eqnarray}
contains all connected diagrams.
This term does not contain unconnected diagrams at all. The other
terms of $n$-th order $E_{0}^{(n) {\rm n} }$ contain unconnected
diagrams modifying the powers of the corresponding electron
propagators in the previous orders. If
the dependence of $E_{0}^{(s) {\rm c} }$$(s < n)$ on $E_{0}^{(0)}$
will be conserved explicitly then
$E_{0}^{(n){\rm n} }$ can be represented as a function of $E_{0}^{(s) {\rm
c} }$ and their derivatives. For example, $E_{0}^{(n) {\rm n} }$ is
written for some particular cases as follows
\begin{eqnarray} \label{noncon}
E_{0}^{(4) {\rm n} } &=& E_{2}\, E_{2}^{\prime},
\\
E_{0}^{(6) {\rm n} } &=& \frac{1}{2!} (E_{2})^{2}\,
E_{2}^{\prime\prime} + E_{2} (E_{2}^{\prime})^{2} +
(E_{2} E_{4})^{\prime},
\\
\label{non8} E_{0}^{(8) {\rm n} } &=&
E_{2} (E_{2}^{\prime})^{3} +3 \frac{1}{2!} (E_{2})^{2} E_{2}^{\prime}
E_{2}^{\prime\prime} + \frac{1}{3!} (E_{2})^{3}
E_{2}^{\prime\prime\prime}
\nonumber \\
&&+
E_{4} (E_{2}^{\prime})^{2} + 2 \frac{1}{2!} E_{2} E_{2}^{\prime\prime}
E_{4} + (E_{2} E_{6})^{\prime} +
E_{4} E_{4}^{\prime}
\nonumber \\
&&+ 2 E_{2} E_{2}^{\prime} E_{4}^{\prime} +
\frac{1}{2!} (E_{2})^{2} E_{4}^{\prime\prime}
,
\end{eqnarray}
where $E_{n}=E_{0}^{(n){\rm c}}$ and a prime denotes
a derivative with respect to $E_{0}^{(0)}$.
All integrals (\ref{rule}) are evaluated analytically by means of
the residue theory \cite{morse} without expanding them in powers of $P$.
Then the effective mass of an electron is defined by the next
formula
\begin{equation}\label{mass}
\displaystyle{ \frac{1}{2 m^{*}} = \left.
\frac{\partial^{2} E_{0}}{\partial P^{2}}\right\vert_{P = 0}. }
\end{equation}

We should like to note that the suggested matrix diagrammatic technique is
acceptable for any $N$-dimensional
optical large polaron, but the rule (\ref{rule})
has to be generalized with respect to the Feynman rules for $N$-
dimensional polaron \cite{smond}.

\section{Results}
Let us carry out an asymptotic expansion of the ground-state energy
up to the eighth order of the PT. At first, it is necessary
to generate and evaluate all connected diagrams for the corresponding
orders with respect to the
conditions (\ref{cond}) and rule (\ref{rule}). At second, we have to
summarize obtained diagrammatic terms and unconnected diagrammatic
terms defined by Eq. (\ref{noncon})-(\ref{non8}). Thus, the polaron
ground-state energy up to the eighth order is
$E_{0} (P) = \sum_{n=0}^{8} E_{0}^{(n)} (P)$,
where the energy terms are defined by
\begin{eqnarray}\label{e0}
E_{0}^{(0)} (P) &=& \frac{P^{2}}{2},
\\
\label{e2} E_{0}^{(2)}  (P) &=&
- \frac{2^{1/2} }{ \left( 2 - P^{2} \right)^{1/2} } \, \alpha=
- \alpha - \frac{P^{2}}{4}\, \alpha + o(P^{4}),
\\
\label{e4} E_{0}^{(4)}  (P) &=&
- \left[
\frac{P^2 (P^2 - 4) + 6 }{(2-P^2)^{3/2} (4-P^2)^{1/2}}
- \frac{P^2 (P^2-3) + 4 }{(2-P^2)^2}
\right]\, \alpha^{2}
\nonumber \\
&=& - \left( \frac{ 3\sqrt{2} }{4} - 1 \right)\, \alpha^{2}
+ \frac{P^{2}}{32} (8 - 5\sqrt{2})\, \alpha^{2} + o(P^{4}) ,
\\
E_{0}^{(6)}  (P) &=& -
\left( 5 - \frac{63}{8\sqrt{2}} + \frac{1}{16}
\sqrt{ \frac{4931}{3} - 1102 \sqrt{2} } \right)\, \alpha^{3}\nonumber \\
&&- \frac{P^{2}}{2} \left( -\frac{15}{4} +
\frac{163}{32\sqrt{2}} + \frac{1}{96}
 \sqrt{ \frac{98593}{6} - 11472 \sqrt{2} } \right) \, \alpha^{3}
+ o(P^{4}),
\\
\label{e8}
E_{0}^{(8)}(P) &=& - \left(
\frac{442369}{15456} - \frac{218861}{7728\sqrt{2}}
+ \frac{151925}{2208\sqrt{3}} - \frac{261335}{2208\sqrt{6}}
\right)
\, \alpha^{4} + o(P^{2}) .
\end{eqnarray}
Since the terms $E_{0}^{(6)}  (P)$ and $E_{0}^{(8)}  (P)$ are too bulky,
we have only written out
their expansion in powers of momentum $P$. Using
Eq. (\ref{e0})-(\ref{e8}) the ground-state energy of a slow-moving
polaron is written as
\begin{eqnarray}\label{asym}
E_{0} (P) & = &\frac{P^{2}}{2 m^{*}} - \alpha
  - 0.06066017 \, \alpha^{2}
\nonumber \\ &
-& 0.00844437 \, \alpha^{3}
- 0.00151488 \, \alpha^{4} + o(\alpha^{5}).
\end{eqnarray}
The effective mass of an electron is defined by equation Eq.
(\ref{mass})
\begin{eqnarray}\label{mass2}
m^{*} &=& 1 + \frac{\alpha}{2} +
\frac{ 5 - 2 \sqrt{2} }{ 8\sqrt{2} }\; \alpha^{2}
 + \left(
- \frac{33}{8} + \frac{183}{32\sqrt{2}}  \right.
\nonumber \\&&
\left. + \frac{1}{2} \sqrt{ \frac{98593}{13824} -
\frac{239}{24 \sqrt{2} }  }
 \right)
\, \alpha^{3} +
o(\alpha^{4}) \simeq 1 + 0.5\, \alpha
\nonumber \\
&& +
0.1919417\, \alpha^{2} + 0.0691096\, \alpha^{3} + o(\alpha^{4}).
\end{eqnarray}

Now let us compare the obtained asymptotic formula for the polaron
ground-state energy with the energy obtained in the frame of
the Feynman polaron theory \cite{fey,deg2} (see Table \ref{tab2}).
For
$\alpha < 3.4$ the asymptotic energy Eq. (\ref{asym}) lies lower than
the Feynman variational
result $E_{0}^{F}$ with maximum deviation about $4$ percent.
For $\alpha \gtrsim 5$
formula Eq. (\ref{asym}) is not correct because of
the radius of convergence of the series is
$R \sim 5$ (see below).
 Note that the first three terms of the energy Eq. (\ref{asym})
coincide with the same terms from Ref.
\cite{ganb2,peetsm}.

Let us estimate the radius of convergence of the PT series
for $E_{0} (0)$.
The radius of convergence $R$ can be evaluated by Cauchy-Hadamar's
criterion \cite{morse}
\begin{eqnarray}
\displaystyle{R = \lim_{n\rightarrow \infty} R_{n} =
\lim_{n\rightarrow \infty} \left( \vert E_{0}^{(n)} \vert
/ \alpha^{n/2} \right)^{-2/n} .}
\end{eqnarray}
It is clear from Table \ref{tab1} that
 there is quite fast convergence
of the sequence $\{ R_{n} \}$ near the point $\alpha \sim 5$.
So if
the nonevaluated higher order energy terms
conserve the existent tendency to convergence of the
sequence $\{ R_{n} \}$ then
the series Eq. (\ref{asym}) has a finite radius of convergence
$R \sim 5$.

\section{Conclusion}
The main purpose of the paper is to develop
the matrix diagrammatic
technique for the optical large
polaron problem in the weak coupling limit.
 The first four terms of the ground-state energy and the first three
terms of the effective mass of the
one-dimensional polaron
are evaluated by means of this technique.
The suggested technique is
acceptable for any $N$-dimensional optical large polaron.
The obtained results are compared with the
results from the Feynman polaron theory.
The radius of convergence of the PT series
for the one-dimensional polaron is estimated by
Cauchy-Hadamar's criterion.

\section*{Acknowledgments}
The author should like to acknowledgement L.~I.~Komarov and I.~D.~Feranchuk
for useful discussions. The author is also grateful to the referee
for constructive criticism.

\begin{table}
\caption{ The ground-state energy $E_{0} (P)$.
}
\label{tab2}
\begin{tabular}{ccc}
$\alpha$  & $-E_{0}^{F}$ & $-E_{0} (0)$ from (\ref{asym})   \\
\hline
$0.1$ & $0.100376$ & $0.100615$  \\
$0.5$ & $0.510063$ & $0.516315$  \\
$1.0$ & $1.044445$ & $1.070619$  \\
$1.5$ & $1.613146$ & $1.672654$  \\
$2.0$ & $2.236957$ & $2.334434$  \\
$2.5$ & $2.959682$ & $3.070245$  \\
$3.0$ & $3.828595$ & $3.896646$\\
$3.3$ & $4.426768$ & $4.443709$\\
$3.4$ & $4.639049$ & $4.635570$\\
$3.5$ & $4.857770$ & $4.832468$\\
$4.0$ & $6.047798$ & $5.898815$\\
$4.5$ & $7.398112$ & $7.119062$\\
$5.0$ & $8.908301$ & $8.518858$\\
\end{tabular}
\end{table}
\begin{table}
\caption{First four terms of the sequence $\{ R_{n}\}$}
\label{tab1}
\begin{tabular}{cccccc}
$n$     & 2 & 4          & 6          & 8         & $\infty$    \\
\hline
$R_{n}$ & 1 & 4.060207   & 4.910708   & 5.068795  & $R$\\
\end{tabular}
\end{table}
\end{document}